\documentclass[12pt]{iopart}
\usepackage{iopams,graphicx}

\usepackage{dcolumn}
\usepackage{bm}
\usepackage{graphicx}
\usepackage{comment}
\usepackage[caption=false]{subfig}
\usepackage{color}

\newcommand{\bea}{\begin{eqnarray}}

\newcommand{\eea}{\end{eqnarray}}

\begin{document}

\title[Power-law relaxation of a confined diffusing particle with memory]{Power-law relaxation of a confined diffusing particle subject to resetting with memory}
\author{Denis Boyer}
\address{Instituto de F\'\i sica, Universidad Nacional Aut\'onoma de M\'exico, 
Ciudad de M\'exico 04510, M\'exico}
\author{Satya N. Majumdar}
\address{LPTMS, CNRS, Univ.  Paris-Sud,  Universit\'e Paris-Saclay,  91405 Orsay,  France}

\begin{abstract}
We study the relaxation of a Brownian particle  with long range memory under confinement in one dimension. The particle diffuses in an arbitrary confining potential and resets at random times to previously visited positions, chosen with a probability proportional to the local time spent there by the particle since the initial time. This model mimics an animal which moves erratically in its home range and returns preferentially to familiar places from time to time, as observed in nature. The steady state density of the position is given by the equilibrium Gibbs-Boltzmann distribution, as in standard diffusion, while the transient part of the density can be obtained through a mapping of the Fokker-Planck equation of the process to a Schr\"odinger eigenvalue problem. Due to memory, the approach at late times toward the steady state is critically self-organised, in the sense that it always follows a sluggish power-law form, in contrast to the exponential decay that characterises Markov processes. The exponent of this power-law depends in a simple way on the resetting rate and on the leading relaxation rate of the Brownian particle in the absence of resetting. We apply these findings to several exactly solvable examples, such as the harmonic, V-shaped and box potentials.
\end{abstract}

\maketitle

\section{Introduction}

Let us consider a single Brownian particle which, in addition to diffusion, undergoes a resetting process with rate $r$ to previously visited positions. Resetting follows a stochastic rule that incorporates memory effects: the particle that resets at some time $t$ chooses uniformly at random a preceding time $0\le t'\le t$, {\it i.e.}, with probability density $1/t$, and occupies the position at which it was located at $t'$ again. Thus the resetting move tends to bring the particle near the most visited positions in the past. This model was introduced in the context of random walks on lattices and solved exactly in Ref.~\cite{BS2014}. The model was further extended to a free Brownian particle on the line \cite{BEM2017}. The position distribution at time $t$ was found to approach a Gaussian form at late times with the variance growing anomalously slowly as $ (2D/r)\, \ln t$ for large $t$, where $D$ is the diffusion constant of the particle. This slow dynamics emerges from the resetting induced memory effect: the particle is sluggish to move away from its familiar territory, where the most visited sites are located. Nevertheless, the position distribution is always time-dependent and does not approach a stationary state at late times, contrary to diffusive processes subject to resetting to a single point, which generically admit non-equilibrium steady states (NESSs) \cite{EM2011,EMS2020,MSS2015}.

Various other generalisations of this simple model have been studied in the recent past, for instance by considering a decaying memory \cite {BR2014}, L\'evy flights \cite{BP2016} or active particles \cite{BM2024}. A central limit theorem and an anomalous large deviation principle have also been established for a class of memory walks of this type, for a broad range of memory kernels~\cite{MU2019,BM2023}. The rigorous proofs of these latter results are based on a connection between the resetting process to previous sites and the growth of weighted random recursive trees.

In a more applied context, random walks with long range memory have become increasingly useful in ecology for the description and analysis of animal mobility. There is mounting evidence that animals do not follow pure Markov processes but use their memory and tend to revisit preferred places during ranging \cite{BDF2008,F2008,VM2009,F2013,MFM2014,F2023}. The model above was able to describe quantitatively the movement patterns of Capuchin monkeys in the wild \cite{BS2014}, as well as of individual elks released in an unknown environment \cite{FC2021}. Another feature shared by many animal trajectories is spatial localisation, by opposition to unbounded diffusion, as home ranges are often clearly identifiable. 
The range-resident behaviour of animals can be modelled by incorporating in the simple random walk a central place (such as a den) toward which the walker is attracted,  producing an effective $|x|$ potential and therefore a stationary 'equilibrium' state at large times \cite{ML2006}. Recent studies have considered harmonic potentials and Ornstein–Uhlenbeck (OU) processes as the starting point of models of home range movements of increasing complexity \cite{F2017,MG2020,F2022}.

In this paper, we study a Brownian particle in one dimension under the combined effects of confinement by an external potential $U(x)$ and of memory, as described in the rule above. In this situation, we expect the position distribution $P_r(x,t)$ to approach a stationary form at late times, where the subscript $r$ in $P_r(x,t)$ denotes the nonzero resetting rate. We first recall that in the absence of resetting, {\it i.e.}, for $r=0$, the system approaches the equilibrium stationary state characterised by the Gibbs-Boltzmann distribution
\begin{equation}
P_0^{\rm st}(x)= \frac{1}{Z}\, e^{-\frac{1}{D}\, U(x)}\, ,
\label{Gibbs.1}
\end{equation}
where the friction coefficient of the particle is set to $1$ from now on and the partition function $Z= \int_{-\infty}^{\infty} e^{-U(x)/D}\, dx$ normalises
the distribution to unity. We assume that $U(x)$ is sufficiently confining so that the integral $Z$ is finite.
Moreover, the position distribution relaxes generically to the equilibrium measure in Eq. (\ref{Gibbs.1}) exponentially
fast. More precisely, writing $P_0(x,t)= P_0^{\rm st}(x) + P_0^{\rm tr}(x,t)$ where the superscript '${\rm tr}$'
denotes the time-dependent transient part of the distribution, one finds that for large $t$
\begin{equation}
P_0^{\rm tr}(x,t) = P_0(x,t)- P_0^{\rm st}(x) \sim e^{-\lambda_1 t}\, ,
\label{trans0.1}
\end{equation}
where $\lambda_1$ denotes the first nonzero eigenvalue of the diffusion operator in the confining potential $U(x)$.
The precise value of $\lambda_1$ depends on the potential $U(x)$. The $x$ dependence of the leading 
transient is implicit in the amplitude on the right hand side (rhs) of Eq. (\ref{trans0.1}). In fact, the decay exponent $\lambda_1$, up to a multiplicative constant, is
precisely the gap between the first excited state $E_1$ and the ground state $E_0=0$
of the associated Schr\"odinger operator \cite{Risken} (see Section II).

Here, we investigate the behaviour of the position distribution $P_r(x,t)$, when one switches memory on with a non-zero rate $r$. In the presence of a finite resetting rate $r>0$, we again expect the position distribution to
approach a stationary form $P_r^{\rm st}(x)= P_r(x, t\to \infty)$ at late times. Our first
goal is to characterise this stationary
position distribution. Moreover, 
it is also important to study how the system relaxes to this stationary state for nonzero $r$, i.e.,
how the result in Eq. (\ref{trans0.1}) gets modified with a nonzero $r$. Our main results
are twofold:

\vskip 0.2cm 
\begin{itemize}

\item [] $(i)$ We show that the stationary position distribution $P_r^{\rm st}(x)$
for nonzero $r$ remains of the Gibbs-Boltzmann form independently of $r$, i.e.,
\begin{equation}
P_r^{\rm st}(x) = \frac{1}{Z}\, e^{-\frac{1}{D}\, U(x)}\, .
\label{Gibbs.2}
\end{equation}
Like in the $r=0$ case, this stationary state for $r>0$ is actually an equilibrium state. This property markedly differs from the well-studied problem of a Brownian particle in a potential $U(x)$ and subject to resetting to a fixed position \cite{P2015,ANBND2019,MBMS2020,GPKP2020,CDG2021,RR2021,ARD2022,AD2023,JBPD2023}, a case that manifestly violates detailed balance and generates a NESS. 

\item [] $(ii)$ Even more interestingly, we find that in complete contrast to the exponential decay in Eq. (\ref{trans0.1}) for $r=0$, the relaxation to the stationary state for $r>0$ is always algebraic, or \lq\lq critical". More
precisely, the analogue of Eq. (\ref{trans0.1}) now reads for large $t$
\begin{equation}
P_r^{\rm tr}(x,t)= P_r(x,t)-P_r^{\rm st}(x) \sim t^{-\theta_1}
\label{transr.1}
\end{equation}
where the dependence of the amplitude on $x$ and $r$ is implicit. The exponent $\theta_1$ depends on $r$ and also on the confining potential $U(x)$. In fact we show that
the exponent $\theta_1$ can be expressed in terms of the decay rate $\lambda_1$ of the resetting-free case via the exact relation
\begin{equation}
\theta_1= \frac{\lambda_1}{r+\lambda_1}\, .
\label{rel.1}
\end{equation}
Thus as $r$ increases, the exponent $\theta_1$ decreases and the relaxation dynamics becomes slower.
\end{itemize}

The rest of the paper is organised as follows. In Section 2, to set the stage and our notations, we recall well-known results and the method to study diffusion in a confining potential in the absence of resetting, i.e., for $r=0$. In Section 3, we derive our main results for the stationary state and the transient part in the presence of a finite resetting rate $r>0$. In Section 4, after recovering previous results for the unbounded case $U(x)=0$, we exemplify the calculations for the harmonic and $|x|$ potentials. Section 5 is devoted to the case of a particle in a box, $x\in [0,L]$ with reflecting boundary conditions at the two boundaries, where we derive the full probability distribution $P_r(x,t)$ at all times. A comparison of the theoretical results with numerical simulations is given in Section 6 and we conclude in Section 7.

\section{Diffusion in a confining potential without resetting}\label{r0}

In this section, we recall the basic results and the general method to compute the time-dependent position
distribution of a  Brownian particle diffusing in the presence of a confining potential $U(x)$~\cite{Risken}.
The position $x(t)$ of the particle evolves via the over-damped Langevin equation
\begin{equation}
\frac{dx}{dt}= - U'(x) + \eta(t)\, ,
\label{lange.1}
\end{equation}
where $\eta(t)$ is a Gaussian white noise with zero mean $\langle \eta(t)\rangle=0$ and a delta correlator
$\langle \eta(t)\eta(t')\rangle= 2\, D\, \delta(t-t')$, with $D$ the diffusion constant. (We recall that the friction coefficient is set to unity.) The associated
Fokker-Planck equation reads
\begin{equation}
\partial_t P_0(x,t)= D\, \partial_x^2 P_0(x,t) + \partial_x\left[ U'(x)\, P_0(x,t)\right]\, ,
\label{fp0.1}
\end{equation}
starting from the initial condition $P_0(x,0)=\delta(x-x_0)$ and satisfying the boundary conditions:
$P_0(x\to \pm \infty, t)=0$. 

To solve this partial differential equation, one can use the method of
separation of variables and use the ansatz
\begin{equation}
P_0(x,t) = \phi(x)\, f(t)\, .
\label{sep0.1}
\end{equation}
Substituting (\ref{sep0.1}) in (\ref{fp0.1}) and dividing both sides by $\phi(x)\, f(t)$ one gets
\begin{equation}
\frac{ {\dot f}(t)}{f(t)}= \frac{ D \phi''(x) + U'(x)\, \phi'(x)+ U''(x)\, \phi(x)}{\phi(x)}= -\lambda\, ,
\label{sep0.2}
\end{equation}
where $\lambda$ is a constant independent of $x$ and $t$. Solving for $f(t)$ gives $f(t)= A_0\, e^{-\lambda t}$
where $A_0$ is a constant and one must have $\lambda\ge 0$ so that the solution does not diverge
as $t\to \infty$. The function $\phi(x)$ satisfies a second order differential equation
\begin{equation}
D\, \phi''(x) + U'(x)\, \phi'(x)+ U''(x)\, \phi(x)= -\lambda\, \phi(x)\, ,
\label{phix0.1}
\end{equation}
where $\lambda$ plays the role of an eigenvalue, yet to be determined. 
One can further reduce this eigenvalue equation to a more familiar Schr\"odinger form via
the substitution 
\begin{equation}
\phi(x) = e^{-\frac{1}{2D}\, U(x)}\, \psi (x)\, .
\label{psix0.1}
\end{equation}
It is then easy to see that $\psi(x)$ satisfies the  time-independent Schr\"odinger equation (setting
$m=\hbar=1$)
\begin{equation}
-\frac{1}{2}\, \psi''(x) + V_Q(x)\, \psi(x)= \frac{\lambda}{2D}\, \psi(x)\, ,
\label{sch0.1}
\end{equation}
where the quantum potential $V_Q(x)$ is expressed in terms of the classical potential $U(x)$ via the relation
\begin{equation}
V_Q(x)= -\frac{1}{4D} U''(x)+ \frac{1}{8 D^2}\, [U'(x)]^2\, .
\label{VQ0.1}
\end{equation}

Thus one needs to solve the Schr\"odinger equation (\ref{sch0.1}), find its spectrum, i.e., the eigenvalues
$E_n= \lambda_n/(2D)$ and the associated eigenfunctions $\psi_{\lambda_n}(x)$. Note that in general, the
spectrum of $V_Q(x)$ may consist of both bound and scattering states, even though $U(x)$ may be fully confining.
Once we have the full spectrum, one can then write down the most general solution of Eq. (\ref{fp0.1})
as a linear combination of these eigenfunctions
\begin{equation}
P_0(x,t) = \sum_{n} a_{\lambda_n}\, \left[e^{-\frac{1}{2D}\, U(x)}\, \psi_{\lambda_n}(x)\right]\, e^{-\lambda_n t}\, ,
\label{decomp0.1}
\end{equation}
where the coefficients 
$\{a_{\lambda_n}\}$ can be determined from the initial condition.
Note that the quantum potential $V_Q(x)$ is such that its ground state corresponds to zero energy $\lambda_0/2D=E_0=0$,
with eigenfunction $\psi_0(x)\propto  e^{-U(x)/{2D}}$ as one can easily check. Thus separating the
ground state from the rest of the spectrum in Eq. (\ref{decomp0.1}), we get
\begin{equation}
P_0(x,t) = a_0\, e^{- \frac{1}{D}\, U(x)} + \sum_{\lambda_n>0} a_{\lambda_n}\, \left[e^{-\frac{1}{2D}\, U(x)}\, \psi_{\lambda_n}(x)\right]\, 
e^{-\lambda_n\, t}\, .
\label{decomp0.2}
\end{equation}
One then immediately identifies the first term on the rhs of (\ref{decomp0.2}) as the stationary Gibbs distribution
\begin{equation}
P_0^{\rm st}(x)= a_0 e^{- \frac{1}{D}\, U(x)}\; \quad {\rm with}\quad a_0=\frac{1}{Z}= \frac{1}{\int_{-\infty}^{\infty} e^{-U(x)/D}\, dx}\, .
\label{Pst0.1}
\end{equation}
The second term in (\ref{decomp0.2}), summing over all nonzero modes, represents the transient part of the distribution
$P_0^{\rm tr}(x,t)$. If the first excited state $\lambda_1>0$ is strictly positive, i.e., well separated
from the ground state $\lambda_0=0$, then it governs the leading late time decay of the transient and one obtains
\begin{equation}
P_0^{\rm tr}(x,t) \approx a_{\lambda_1} \,e^{-\frac{1}{2D}\, U(x)}\, \psi_{\lambda_1}(x)\,  e^{-\lambda_1\, t}\, .
\label{Ptr0.1}
\end{equation}
Note that the result in Eq. (\ref{Ptr0.1}) holds provided that $a_{\lambda_1}\ne 0$.
Sometimes, the initial condition may have a special symmetry that renders $a_{\lambda_1}=0$. 
In that case, the leading decay
will be governed by the next nonzero excited state.

Let us remark that the result in Eq. (\ref{Ptr0.1}) is valid for any confining potential $U(x)$, such
that $(i)$ the integral $Z=\int_{-\infty}^{\infty} e^{-U(x)/D}\, dx$ exists, $(ii)$ the associated 
quantum potential $V_Q(x)$ in Eq. (\ref{VQ0.1}) is such that its first excited state
has eigenvalue $\lambda_1>0$ (recall that the ground state has zero energy if a steady state exists).
In addition, the leading decay in Eq. (\ref{Ptr0.1}) will be a pure exponential if $\lambda_1<\lambda_2$, i.e., there is a nonzero gap between the first and the second excited state.
There are some potentials $U(x)$ for which $\lambda_1>0$ may be gapped from the ground state, but
then there is a continuum of scattering states beyond $\lambda_1$. In this case, depending on how the density
of scattering states vanishes as one approaches $\lambda_1$ from above, one may have an additional
sub-leading algebraic correction to the leading exponential decay $\sim e^{-\lambda_1\, t}$. We discuss below
two examples that can be explicitly solved to illustrate these features.

\vskip 0.3cm

\noindent {\bf  Ornstein-Uhlenbeck (OU) process.} As a first simple example, consider the OU process with stiffness $\mu>0$, for which $U(x)= \mu x^2/2$. In this 
case, the quantum potential in Eq. (\ref{VQ0.1}) reads
\begin{equation}
V_Q(x)=\frac{\mu^2}{8 D^2}\, x^2 - \frac{\mu}{4D}\, .
\label{ho.1}
\end{equation}
Thus, one has a quantum harmonic oscillator of mass $m=1$ and frequency $\omega=\mu/(2D)$, with the height of the potential shifted
by $\mu/(4D)$. Consequently, the energy levels are well separated from each other and are given by
\begin{equation}
E_n= \left(n+\frac{1}{2}\right)\, \omega- \frac{\mu}{4D}= n\, \frac{\mu}{2D}\, , \quad {\rm with}\quad n=0,1,2\ldots
\label{ho_spec.1}
\end{equation}
The eigenvalues are given by $\lambda_n= 2D E_n= n\, \mu$ and the leading transient in Eq. (\ref{Ptr0.1})
decays as 
\begin{equation}
P_0^{\rm tr}(x,t) \approx a_{\lambda_1} \,e^{-\frac{\mu}{2D}\, x^2 }\, \psi_{\lambda_1}(x)\,  e^{-\mu\, t}\, ,
\label{Ptrho.2}
\end{equation}
where $\psi_1(x)$ represents the eigenfunction associated with the first excited state of the harmonic oscillator.
Note that if the initial condition is symmetric, e.g., when $P_0(x)=\delta(x)$, this symmetry is preserved at all times, indicating that
only the even eigenfunctions contribute to the spectral decomposition in Eq. (\ref{decomp0.1}). This means that all the odd coefficients vanish:
$a_{\lambda_{2m+1}}=0$ for all $m=0,1,2,\ldots$. In that case, the leading transient corresponds to the second excited
state of the harmonic oscillator and $P_0^{\rm tr}(x,t)$ decays as $\sim e^{-2 \mu t}$ at late times.

\vskip 0.3cm

\noindent {\bf  The potential $U(x)= F\, |x|$ with $F>0$.} This is an interesting example since, in this case, the quantum potential in Eq. (\ref{VQ0.1})
reads
\begin{equation}
V_Q(x)= -\frac{F}{2D}\, \delta(x) + \frac{F^2}{8 D^2}\, .
\label{modx.1}
\end{equation}
This quantum potential has a single bound state at $E=0$ and a continuum of excited states with energies $E\ge E_1=F^2/(8D^2)$.
Thus there is a gap $\lambda_1= 2D\, E_1= F^2/(4D)$, and we expect that the transient in Eq. (\ref{Ptr0.1})
will decay, to leading order for large $t$, exponentially as $e^{-\lambda_1 t}\sim e^{- F^2 t/(4D)}$.
However, unlike in the harmonic oscillator example above, the eigenvalue $\lambda_1$ is not isolated
here and there is a continuum of scattering states above $E_1$. Moreover, the density of such states vanishes
as $\sqrt{E- E_1}$ as $E\to E_1$ from above (see Section \ref{harmlin}) and the leading decay of the transient in Eq. (\ref{Ptr0.1})
now behaves as $e^{- F^2t/(4D) }\, t^{-3/2}$. The sub-leading $t^{-3/2}$ decay emerges from replacing the sum
in Eq. (\ref{decomp0.2}) by an integral above $E_1$. Since the density of states vanishes as $\sqrt{E- E_1}$, this
leads to the power law correction $t^{-3/2}$ multiplying the leading exponential $e^{- F^2t/(4D) }$~\cite{SM2020}.

\section{Diffusion in a confining potential in the presence of resetting}

We now allow the particle to perform resetting moves, in addition to the diffusion in the confining external potential.
Let us recall the memory-driven resetting dynamics. At time $t$, with rate $r$ the particle chooses any previous time in the past $0\le t'< t$ with
uniform probability density $1/t$ and resets to the position that it occupied at that previous instant $t'$. Let $P_r(x,t)$
denote the position distribution at time $t$. To take into account the memory effect, we now need to define the
two-point function $P_r(x,t\,; x', t')$, which is the joint probability density for the particle to be near $x'$ at $t'$
and near $x$ at $t$, with $t'\le t$. Clearly, if we integrate over one of the positions, we recover the
marginal one-point probability density
\begin{equation}
\fl \int_{-\infty}^{\infty} P_r(x,t\, ; x', t')\, dx'= P_r(x,t) \quad {\rm and}\quad \int_{-\infty}^{\infty}P_r(x,t\, ; x', t')\, dx= P_r(x',t').
\label{marg.1}
\end{equation}
With these ingredients at hand, we can now write down the Fokker-Planck equation for the evolution of the one-point
position distribution $P_r(x,t)$ as
\begin{eqnarray}
\partial_t P_r(x,t)&=& D\, \partial_x^2 P_r(x,t) + \partial_x\left[ U'(x)\, P_r(x,t)\right] - r\, P_r(x,t)\nonumber\\ 
&&+
\frac{r}{t}\, \int_0^t dt' \int_{-\infty}^{\infty} dx'\, P_r(x',t; x, t')\, .
\label{fpr.1}
\end{eqnarray}
The first two terms on the rhs describe, as before, the diffusion in the external potential $U(x)$. The third term
describes the loss of probability density from position $x$ at time $t$ due to resetting to other positions.
The last term on the rhs describes the gain in the probability density at $x$ at time $t$ due to resetting
from other positions labelled by $x'$. If the particle has to reset to $x$ from $x'$ at time $t$, it must have been
at $x$ at a previous time $t'\le t$ and the probability density of this event is simply the
two-point function $ P_r(x',t; x, t')$. Finally, we need to integrate over all positions $x'$ from
which the particle may arrive at $x$ by resetting. The reason for the solvability for the one-point
position distribution in this model can then traced back to the second relation in Eq. (\ref{marg.1}), which
allows us to write a closed equation for the one-point function (without involving multiple-point functions) provided that the resetting rate $r$ is independent of the position. We obtain
\begin{equation}
\fl \partial_t P_r(x,t)= D\, \partial_x^2 P_r(x,t) + \partial_x\left[ U'(x)\, P_r(x,t)\right] - r\, P_r(x,t) 
+\frac{r}{t}\, \int_0^t dt' \,  P_r(x, t')\, .
\label{fpr.2}
\end{equation}
It is easy to check that the total probability $\int_{-\infty}^{\infty} P_r(x,t)\, dx$ remains conserved and equals unity due to the initial condition, e.g., $P_r(x,0)= \delta(x-x_0)$. One also needs to impose the vanishing boundary conditions as $x\to \pm \infty$, i.e.,
$P_r(x\to \pm \infty, t)=0$ for all $t$. Evidently, for $r=0$, Eq. (\ref{fpr.2}) reduces to the standard
Fokker-Planck equation (\ref{fp0.1}).

To solve the Fokker-Planck equation (\ref{fpr.2}), we follow the same route as in the $r=0$ case, namely, we use the
method of separation of variables with the ansatz
\begin{equation}
P_r(x,t) = \phi(x)\, f_r(t)\, .
\label{sepr.1}
\end{equation}
Substituting (\ref{sepr.1}) in Eq. (\ref{fpr.2}), dividing both sides by $\phi(x) f_r(t)$ and assembling
the only-time dependent and only-space dependent parts separately gives
\begin{equation}
\fl\frac{{\dot f_r}(t)}{f_r(t)} + r - \frac{r}{t\, f_r(t)}\, \int_0^t f_r(t')\, dt' = \frac{ D \phi''(x) + U'(x)\, \phi'(x)+ U''(x)\, \phi(x)}{\phi(x)}= -\lambda\, ,
\label{sepr.2}
\end{equation}
where $\lambda\ge 0$ is a constant independent of $x$ and $t$. Note that the space-dependent function $\phi(x)$ is independent of $r$
and satisfies the same eigenvalue equation (\ref{phix0.1}) as the $r=0$ case. Consequently, the analysis of the spatial part
is exactly as in the $r=0$ case discussed in the previous section. In particular, we get
\begin{equation}
\phi(x)= e^{-\frac{1}{2D}\, U(x)}\, \psi (x)\, ,
\label{phixr.1}
\end{equation}
where $\psi(x)$ satisfies the Schr\"odinger equation (\ref{sch0.1}), with the same eigenvalues
$\lambda_n$'s as in the $r=0$ case.

Thus, the $r$-dependence enters only in the time-dependent part $f_r(t)$ of the solution. 
For a given $\lambda$, we get from Eq. (\ref{sepr.2})
a second order ordinary differential equation for $f_r(t)$
\begin{equation}
{\dot f_r}(t) + (r+\lambda)\, f_r(t)- \frac{r}{t}\, \int_0^t f_r(t')\, dt'=0\, .
\label{ft_diff.1}
\end{equation}
Multiplying this equation by $t$ and taking the time derivative, one obtains
\begin{equation}
t{\ddot f_r}(t) + [1+(r+\lambda)t]\, \dot{f_r}(t)+ \lambda f_r(t)=0\, .
\label{cumul_Ft.1}
\end{equation}
Through the change of variable $z=-(\lambda+r)t$ with $g(z)\equiv f(t)$, Eq. (\ref{cumul_Ft.1}) can be recast as a confluent hypergeometric equation,
\begin{equation}
z\frac{d^2 g(z)}{dz^2}+(b-z)\frac{dg(z)}{dz}-ag(z)=0,
\label{Ft_diff.2}
\end{equation}
with $b=1$ and $a=\lambda/(\lambda+r)$. The differential equation (\ref{Ft_diff.2}) has two linearly independent solutions denoted by $M(a,b,z)$ and $U(a,b,z)$ \cite{AS_book}.
Hence, the most general solution of Eq. (\ref{cumul_Ft.1}) can be expressed as
\begin{equation}
f_r(t) = c_1\,  M \left( \frac{\lambda}{r+\lambda}, 1, - (r+\lambda) t\right)
+ c_2\,\, U \left( \frac{\lambda}{r+\lambda}, 1, - (r+\lambda) t\right)\, ,
\label{gen_sol.1}
\end{equation}
where $c_1$ and $c_2$ are two arbitrary constants. The solution must be finite at all $t$, in particular at $t=0$. The small argument asymptotics of the two solutions are
given by~\cite{AS_book}
\begin{eqnarray}
M(a,b,z) &\to & 1\, \quad\quad\quad {\rm as}\quad z\to 0 \label{Mz0.1} \\
U(a,1,z) &\to & -\frac{\ln z}{\Gamma(a)} \, \quad {\rm as}\quad z\to 0 \, . 
\label{Uz0.1} 
\end{eqnarray}
Consequently, the divergence as $t\to 0$ in Eq. (\ref{gen_sol.1}) is avoided by setting $c_2=0$, and our solution simply reads
\begin{equation}
f_r(t) = M \left( \frac{\lambda}{r+\lambda}, 1, - (r+\lambda) t\right)\, .
\label{ft_sol.1}
\end{equation} 
where we have set $c_1=1$ by absorbing this constant into the amplitude of
the eigenmode. Importantly, using the identity $M(0,b,z)=1$, we recover $f_r(t)=1$ for $\lambda=0$. This means that the solution $\phi(x)$ with $\lambda=0$ in the eigenvalue equation (\ref{sepr.2}), i.e., the Gibbs-Boltzmann distribution $\phi(x)=a_0 e^{-\frac{1}{D}U(x)}$, is also a stationary state when $r>0$. 

To obtain the asymptotic behaviour of $f_r(t)$ in time for $\lambda>0$, it is useful to recall here some
basic properties of the confluent hypergeometric function $M(a,b,z)\equiv {_1}F_1(a,b,z)$. It has a simple 
power series expansion
\begin{equation}
M(a,b,z)= 1 + \frac{a}{b}\, z+ \frac{a(a+1)}{b(b+1)}\, \frac{z^2}{2!}+ \frac{a(a+1)(a+2)}{b(b+1)(b+2)}\, \frac{z^3}{3!}+ \ldots\, ,
\label{M_def.1}
\end{equation}
whereas for large negative arguments, we have~\cite{AS_book}
\begin{equation}
M(a,b,-z) \simeq \frac{\Gamma(b)}{\Gamma(b-a)}\, z^{-a}\,  \quad {\rm as}\quad z\to \infty\, .
\label{M_asymp.1}
\end{equation}
Consequently, $f_r(t)$ in Eq. (\ref{ft_sol.1}) behaves for large $t$ as a power-law,
\begin{equation}
f_r(t) \simeq \frac{(r+\lambda)^{-\frac{\lambda}{r+\lambda}}}{\Gamma\left(\frac{r}{r+\lambda}\right)}\, t^{- \frac{\lambda}{r+\lambda}}\, .
\label{ft_asymp.1}
\end{equation}  
Note that this result does not hold for $r=0$, as the prefactor in Eq. (\ref{ft_asymp.1}) vanishes in this case. Rather, the exponential decay $f_r(t)=e^{-\lambda t} $ is recovered by setting $r=0$ directly in Eq. (\ref{ft_sol.1}), since $M(1,1,-z)=e^{-z}$.

Finally, combining the time-dependent part $f_r(t)$ and the space-dependent part that involves the set of
all eigenvalues of the Schr\"odinger operator, we can express the full solution of $P_r(x,t)$ as
the following exact spectral decomposition
\begin{equation}
P_r(x,t)= e^{-\frac{1}{2D}\, U(x)}\sum_{n} a_{\lambda_n}\, \psi_{\lambda_n}(x)\, 
 M \left( \frac{\lambda_n}{r+\lambda_n}, 1, - (r+\lambda_n) t\right)\, .
\label{Prxt_sol.1}
\end{equation}
As in the $r=0$ case before, we will now
separate out the stationary mode corresponding to the ground state $n=0$ with $\lambda_0=0$ from the transient part
and rewrite Eq. (\ref{Prxt_sol.1}) as
\begin{equation}
\fl P_r(x,t)= a_0\, e^{-\frac{1}{D}\, U(x)} + e^{-\frac{1}{2D}\, U(x)}\sum_{n>0} a_{\lambda_n}\, \psi_{\lambda_n}(x)\,
 M \left( \frac{\lambda_n}{r+\lambda_n}, 1, - (r+\lambda_n) t\right) .
\label{Prxt_sol.2}
\end{equation}
At late times, using the asymptotic behaviour of $f_r(t)$ in Eq. (\ref{ft_asymp.1}), it follows
that the $n$-th mode in the second term on the rhs decays at late times as a power law $\sim t^{-\frac{\lambda_n}{r
+\lambda_n}}$ for $\lambda_n>0$. Consequently, the sum in Eq. (\ref{Prxt_sol.2})
represents the transient part $P_r^{\rm tr}(x,t)$, while the first term that is independent of $t$
represents the stationary solution. 

Hence, we come to the two main results announced in
the Introduction. First, the stationary solution
\begin{equation}
P_r^{\rm st}(x)= a_0 e^{- \frac{1}{D}\, U(x)}\; \quad {\rm with}\quad 
a_0=\frac{1}{Z}= \frac{1}{\int_{-\infty}^{\infty} e^{-U(x)/D}\, dx}\, ,
\label{Pstr.1}
\end{equation}
is independent of $r$ and remains with the Gibbs-Boltzmann form as in the $r=0$ case. This solution corresponds to an equilibrium state, since local detailed balance is fulfilled even with $r>0$. This can be understood by considering two small disjoint regions $[x,x+\delta_x]$ and $[y,y+\delta_y]$, which have been occupied until time $t$ for an total amount of time $\tau_x(t)$ and $\tau_y(t)$, respectively. The probability that the particle takes a move from the region $x$ to $y$ by resetting at time $t$ is $\delta_x P^{\rm st}(x)w^{(r)}_{x\to y}$, where the transition rate is $w^{(r)}_{x\to y}=r\tau_y(t)/t$ due to the preferential revisit rule. Since $\delta_x P^{\rm st}(x)\simeq \tau_x(t)/t$ at large $t$, the above transition probability reads $r\tau_y(t)\tau_x(t)/t^2$, which is symmetric with respect to $x$ and $y$, implying detailed balance. This property is a specificity of the linear reinforcement rule. 
We note that it also follows from Eq. (\ref{fpr.2}): if $P_r(x,t)$ becomes independent of time for all $x$ at large $t$, then the 4th term in the rhs becomes $rP^{\rm st}(x)$ and  cancels with the 3rd term. Therefore the equation for the stationary state is the same as with $r=0$.

Secondly, the leading behaviour of the transient part decays as
\begin{equation}
P_r^{\rm tr}(x,t) \sim e^{-\frac{1}{2D}\, U(x)}\, \psi_{\lambda_1}(x)\, t^{-\theta_1}\, , \quad {\rm with}
\quad \theta_1= \frac{\lambda_1}{r+\lambda_1}\, ,
\label{Pr_tr.1}
\end{equation}
where $\lambda_1=\min_{n\in{\mathbb N}^+}\{\lambda_n\}$ represents the gap between the ground state and the first excited state of the underlying Schr\"odinger operator of the $r=0$ case. Thus, unlike the stationary solution
that remains unaffected by resetting, the transient part gets modified drastically if $r>0$. It now decays very slowly as a power law at late times
with a non-universal exponent $\theta_1= \lambda_1/(r+\lambda_1)$ comprised in the interval $(0,1)$ and that decreases with increasing $r$, indicating that the dynamics get slower
and slower due to resetting. This result is completely general, valid for any shape of the confining potential.

\section{Examples of potentials}\label{harmlin}

\subsection{Unbounded case $U(x)=0$}\label{u=0}

Before proceeding to examples of confining potentials, we show how the free particle results, in particular the logarithmic growth of the mean square displacement \cite{BS2014,BEM2017}, can be recovered from the analysis above. If the particle diffuses in an unbounded space, or $U(x)=0$ for $x\in(-\infty,\infty)$, the Schr\"odinger equation (\ref{sch0.1}) reads $\psi''(x)=-(\lambda/D)\psi(x)$. The non-diverging solutions are of the form $\psi(x)=\cos(kx+\varphi)$ with $\lambda=\lambda_k=Dk^2$ and $k$ real. Since the potential does not have a finite $Z$, $a_0=0$ and Eq. (\ref{Prxt_sol.2}) becomes a continuous sum over the modes $k$
\begin{equation}
 P_r(x,t)= \int_{-\infty}^{\infty} dk\, a_{k}\, \cos(kx+\varphi_k)\,
 M \left( \frac{\lambda_k}{r+\lambda_k}, 1, - (r+\lambda_k) t\right) ,
\label{Prxt_sol.U=0}
\end{equation}
with $a_k$ and $\varphi_k$ determined from the initial condition.
Eq. (\ref{Prxt_sol.U=0})  is equivalent to the solution presented in \cite{BEM2017} under a slightly different form. At late times, with the asymptotics (\ref{M_asymp.1}) the solution simplifies to
\begin{eqnarray}
 P_r(x,t)&&\simeq \int_{-\infty}^{\infty} dk\, \frac{a_k}{\Gamma\left(\frac{r}{r+\lambda_k}\right)}\,\, \cos(kx+\varphi_k)\, [(r+\lambda_k)t]^{-\frac{\lambda_k}{r+\lambda_k}}\\
 &&\simeq\int_{-\infty}^{\infty} dk\, a_{0}\, \cos(kx+\varphi_0)\, e^{-(Dk^2/r)\ln (rt)},\label{asU=0}
\end{eqnarray}
where $r+\lambda_k$ has been replaced by $r$ in the last integral due to the fact that
at large $\ln (rt)$, the small $k$ regime dominates the integration. Expression (\ref{asU=0}) is the inverse Fourier transform of $e^{-\sigma(t)k^2/2}$, therefore the asymptotic form of $P_r(x,t)$ is Gaussian with variance $\sigma(t)=(2D/r)\ln (rt)$.

\subsection{Harmonic potential}

A specific example of confining potential corresponds to the OU process where $U(x)= \mu x^2/2$. In this case,
the quantum potential $V_Q(x)$ represents a harmonic oscillator with
energy eigenvalues $E_n=n\, \frac{\mu}{2D}$ or $\lambda_n= n\,\mu$ with
$n=0,1,2\ldots$, see Eqs. (\ref{ho.1})-(\ref{ho_spec.1}). Consequently, $\lambda_1= \mu$ and the exponent $\theta_1$ in Eq. (\ref{Pr_tr.1})
is given by
\begin{equation}
\theta_1= \frac{\mu}{r+\mu}\, .
\label{OU_exp.1}
\end{equation}

\subsection{V-shaped potential}
For the case $U(x)= F\, |x|$, it turns out that the transient part of the density behaves as
\begin{equation}
P_r^{\rm tr}(x,t) \sim (\ln t)^{-3/2}\, t^{-\theta_1}\, , \quad {\rm with}\quad \theta_1= \frac{\lambda_1}{r+\lambda_1}= 
\frac{F^2}{ 4 D r+F^2}\, . 
\label{modxr.1}
\end{equation}
As discussed in Section \ref{r0}, the first excited
state $\lambda_1= F^2/(4D)$ is well separated from the ground state $E_0=0$, but there is
a continuum of excited states above $E_1=\lambda_1/(2D)$
whose density vanishes as
$\sqrt{E-E_1}$ as $E\to E_1$, as shown below. This leads to additional logarithmic corrections
to the leading power law decay in Eq. (\ref{Pr_tr.1}). 

Let us now derive this result. We recall that the excited eigenfunctions with energy $E>E_1$ of the Schr\"odinger equation with the potential given by Eq. (\ref{modx.1}) are spatially extended and are of two types, symmetric and anti-symmetric, or even and odd in $x$, respectively \cite{Risken}:
\begin{eqnarray}
\psi_{k}^{(s)}(x)&=&\frac{1}{\sqrt{(4k^2+\kappa^2)\pi}}[2k\cos(kx)-\kappa\sin(k|x|)]\label{sym}\\
\psi_{k}^{(a)}(x)&=&\frac{1}{\sqrt{\pi}}\sin(kx)\label{antisym}
\end{eqnarray}
with
\begin{eqnarray}
k&=&\sqrt{(\lambda-\lambda_1)/D}\ge0\label{klambda}\\ 
\kappa&=&F/D, \label{kappa}
\end{eqnarray}
and $\lambda=2DE$ is taken as a continuous parameter.
These functions are $\delta$-normalised, or $\int_{-\infty}^{\infty} dk\, \psi_{k}(x)\psi_{k'}(x)=\delta(k-k')$, therefore it is convenient to use the index $k$ and expand the transient part of the density in this basis. Eq. (\ref{Prxt_sol.2}) becomes
\begin{eqnarray}
&&P_r(x,t)= a_0\, e^{-\frac{1}{D}\, U(x)} \label{Prxt_sol.3}\\
&&+ e^{-\frac{1}{2D}\, U(x)}\int_{0}^{\infty}dk\, \left(\sum_{i=s,a}\alpha_{k}^{(i)}\, \psi_{k}^{(i)}(x)\right)\,
 M \left( \frac{\lambda_k}{r+\lambda_k}, 1, - (r+\lambda_k) t\right)\, ,\nonumber
\end{eqnarray}
where the dispersion relation is given from Eq. (\ref{klambda}) by 
\begin{equation}
\lambda_k=\lambda_1+Dk^2\, .
\label{disp}
\end{equation}
The coefficients $\alpha_k^{(i)}$ follow from the initial condition $P_r(x,t)=\delta(x-x_0)$, where the Kummer's function $M$ in Eq. (\ref{Prxt_sol.3}) is simply unity. By projecting, one obtains
\begin{equation}
\alpha_k^{(i)}=e^{\frac{U(x_0)}{2D}}\psi_{k}^{(i)}(x_0)  ,\quad i=s,a.
\end{equation}
Therefore,
\begin{eqnarray}
&&P_r(x,t)= a_0\, e^{-\frac{1}{D}\, U(x)} \label{Prxt_sol.4}\\
&&+ e^{\frac{U(x_0)-U(x)}{2D}}\int_{0}^{\infty}dk\, \left(\sum_{i=s,a}\psi_{k}^{(i)}(x_0)\, \psi_{k}^{(i)}(x)\right)\,
 M \left( \frac{\lambda_k}{r+\lambda_k}, 1, - (r+\lambda_k) t\right).
\nonumber
\end{eqnarray}
We next expand the integrand of Eq. (\ref{Prxt_sol.4}) at small $k$, as it dominates in the large $t$ limit. For a fixed $x$, we have from Eqs, (\ref{sym})-(\ref{antisym}),
\begin{eqnarray}
\psi_k^{(s)}(x)&\simeq& \frac{1}{\kappa\sqrt{\pi}} k(2-\kappa|x|)\\
\psi_k^{(a)}(x)&\simeq&\frac{1}{\sqrt{\pi}}kx,
\end{eqnarray}
while
\begin{equation}
\frac{\lambda_k}{r+\lambda_k}\simeq 
\frac{\lambda_1}{r+\lambda_1}+\frac{Dr}{(r+\lambda_1)^2}k^2.
\end{equation}
Inserting these expressions into Eq. (\ref{Prxt_sol.4}) and using the large time expressions in Eqs. (\ref{M_asymp.1})-(\ref{ft_asymp.1}) for the Kummer's function, one obtains the leading behaviour in $k$ of the integrand. One notices that when $\ln t$ is large, the Fourier integral is actually controlled by the small $k$ regime,
\begin{eqnarray}
 && P_r(x,t)\simeq a_0\, e^{-\frac{1}{D}\, U(x)}  \label{modxr.1b}\\
 &&\fl+e^{\frac{U(x_0)-U(x)}{2D}}\left[xx_0+\frac{(2-\kappa|x|)(2-\kappa|x_0|)}{\kappa^2}\right]\frac{[(r+\lambda_1)t]^{-\frac{\lambda_1}{r+\lambda_1}}}{\pi\Gamma\left(\frac{r}{r+\lambda_1}\right)}\,\,\int_0^{\infty}
 dk\,k^2\, e^{-\frac{Dr\ln[(\lambda_1+r) t]}{(r+\lambda_1)^2}k^2}.
 \nonumber
\end{eqnarray}
This expansion cannot be used in the limit $x\rightarrow\infty$, as we have considered $kx\ll 1$ or $|x|\ll (\ln t)^{1/2}$. Performing the Gaussian integral and using Eq. (\ref{kappa}) leads to the final expression
\begin{eqnarray}
&&P_r(x,t)\simeq a_0\, e^{-\frac{F|x|}{D}} \label{modxr.2}\\
&&+e^{\frac{F(|x_0|-|x|)}{2D}}\left[xx_0+\frac{(2D-F|x|)(2D-F|x_0|)}{F^2}\right]\frac{(r+\lambda_1)^{\frac{3r+2\lambda_1}{r+\lambda_1}}}{4\sqrt{\pi}\Gamma\left(\frac{r}{r+\lambda_1}\right)}\nonumber\\
&&\quad \times \left( Dr\ln[(r+\lambda_1)t]\right)^{-3/2}\, t^{-\frac{\lambda_1}{r+\lambda_1}},\nonumber
\end{eqnarray}
which is of the form of Eq. (\ref{modxr.1}) at large $\ln t$. From Eq. (\ref{modxr.1b}), we see that the number of states in $[k,k+dk]$ is $k^2dk$ at small wave-numbers, which, from Eq. (\ref{disp}), is equivalent to a density of state going as $\sqrt{E-E_1}$ near $E_1$

\section{Exact solution for a particle in an interval with reflecting boundary conditions}

In this section, we present the exact time-dependent solution for the position distribution
of the particle with memory diffusing in a box potential, i.e., an interval $[0,L]$ with reflecting boundary conditions. This is the case of a singular potential $U(x)$
in Eq. (\ref{fpr.2}). The general mapping to the quantum problem will still work, with
$V_Q(x)$ representing a box potential which is zero in $[0,L]$ and infinite outside. 
However, we can avoid using the quantum mapping and solve directly the Fokker-Planck equation instead. 
In this case, the Fokker-Planck equation (\ref{fpr.2}) reduces to   
\begin{equation}
\partial_t P_r(x,t)= D\, \partial_x^2 P_r(x,t) -r\, P_r(x,t)+ \frac{r}{t}\, \int_0^t dt' \, P_r(x, t'),
\label{fpr_box.1}
\end{equation}
for $0\le x\le L$.
The initial condition is $P_r(x,0)=\delta(x-x_0)$ and $P_r(x,t)$ satisfies
the reflecting boundary conditions at $x=0$ and $x=L$,
\begin{equation}
\partial_x P_r(x,t)\Big|_{x=0}=0 \, \quad {\rm and}\quad \partial_x P_r(x,t)\Big|_{x=L}=0\, .
\label{rf_bc.1}
\end{equation}
We decompose $P_r(x,t)$ into its Fourier modes and write
\begin{equation}
P_r(x,t) = \sum_{k} \tilde{P}_r(k,t)\, \cos(k x)\, .
\label{cosine_decomp.1}
\end{equation}
The reflecting boundary condition at $x=0$ is automatically satisfied since we have only kept the
cosine modes. The boundary condition at $x=L$ imposes $\sin(k L)=0$, implying that
$k$ takes the values $k_n= n\pi/L$ where $n=0,1,2,\ldots$. 
Substituting (\ref{cosine_decomp.1}) into the linear equation (\ref{fpr_box.1}), one finds, as expected,
that the different modes decouple and $\tilde{P}_r(k,t)$ satisfies the evolution equation for any $k=k_n$
\begin{equation}
\frac{d \tilde{P}_r(k,t)}{dt}= - (r+ D\, k^2)\, \tilde{P}_r(k,t) + \frac{r}{t}\, \int_0^t \tilde{P}_r(k,t')\, dt'\, .
\label{pkt_evol.1}
\end{equation}
This equation is the same as Eq. (\ref{ft_diff.1}) with $\lambda=Dk^2$.
Hence, we can directly write down the full exact solution
for $\tilde{P}_r(k,t)$ by replacing $\lambda$ by $Dk^2$ in Eq. (\ref{ft_sol.1}).
Substituting these solutions back into the Fourier decomposition in Eq. (\ref{cosine_decomp.1}) gives
\begin{equation}
P_r(x,t) = \sum_{n=0}^{\infty} a_n\, \cos(k_n x)\,
 M \left( \frac{Dk_n^2}{r+Dk_n^2}, 1, - (r+Dk_n^2)\, t\right) .
\label{box_sol.1}
\end{equation}
The coefficients $a_n$'s can be determined from the initial condition $P_r(x,0)= \delta(x-x_0)$.
At $t=0$, the $M$-functions on the rhs in Eq. (\ref{box_sol.1}) are exactly
$1$. Hence,
\begin{equation}
P_r(x,0)= \delta(x-x_0)= \sum_{n=0}^{\infty} a_n\, \cos\left(\frac{n \pi x}{L}\right)\, .
\label{init.1}
\end{equation}
The orthogonality property of the cosine modes, 
\begin{equation}
\int_0^L \cos\left(\frac{m \pi x}{L}\right)\, \cos\left(\frac{n \pi x}{L}\right)\, dx= \frac{L}{2}\, 
\delta_{m,n}\, ,
\label{ortho.1}
\end{equation}
where $\delta_{m,n}$ is the Kronecker delta function, and $m\ne 0$, $n\ne 0$, gives from Eq. (\ref{init.1})
\begin{equation}
a_n =  \frac{2}{L} \cos \left( \frac{n\pi x_0}{L}\right) \, , \quad {\rm for}\quad n\ge 1 \, 
\label{coeff.1}
\end{equation}
while $a_0=\frac{1}{L}$.
The full exact solution thus reads
\begin{eqnarray}
\fl &&P_r(x,t) = \frac{1}{L}
  \label{box_sol.2}\\ \fl && +\frac{2}{L}\, \sum_{n=1}^{\infty} 
\cos\left( \frac{n\pi x_0}{L}\right)\, \cos\left( \frac{n\pi x}{L}\right) 
 M \left( \frac{D (\frac{n\pi}{L})^2}{r+D (\frac{n\pi}{L})^2}, 1, - \left[r+D\left(\frac{n\pi}{L}\right)^2\right]\, t\right).
  \nonumber
\end{eqnarray}
The first term $1/L$ represents the equilibrium steady state $P_r^{\rm st}(x)$ where the particle gets uniformly distributed
over the full interval, independently of the resetting rate $r$. The second term, the sum over the excited modes, represents the transient part $P_r^{\rm tr}(x,t)$. To leading order for large $t$,
the dominant contribution comes from the mode $n=1$ and  
again decays algebraically as
\begin{equation}
P_r^{\rm tr}(x,t) \sim t^{-\theta_1}\, \quad {\rm with} \quad \theta_1=\frac{ D\,\pi^2/L^2}{r+ D \pi^2/L^2}\, .
\label{box_trans.1}
\end{equation}  
This result is perfectly consistent with the general result in Eq. (\ref{Pr_tr.1}) since
$\lambda_1= D \pi^2/L^2$ indeed represents the gap between the first excited state and the ground state
of a quantum particle in a box $[0,L]$.

\begin{figure}
\centering
\includegraphics[width=0.8\textwidth]{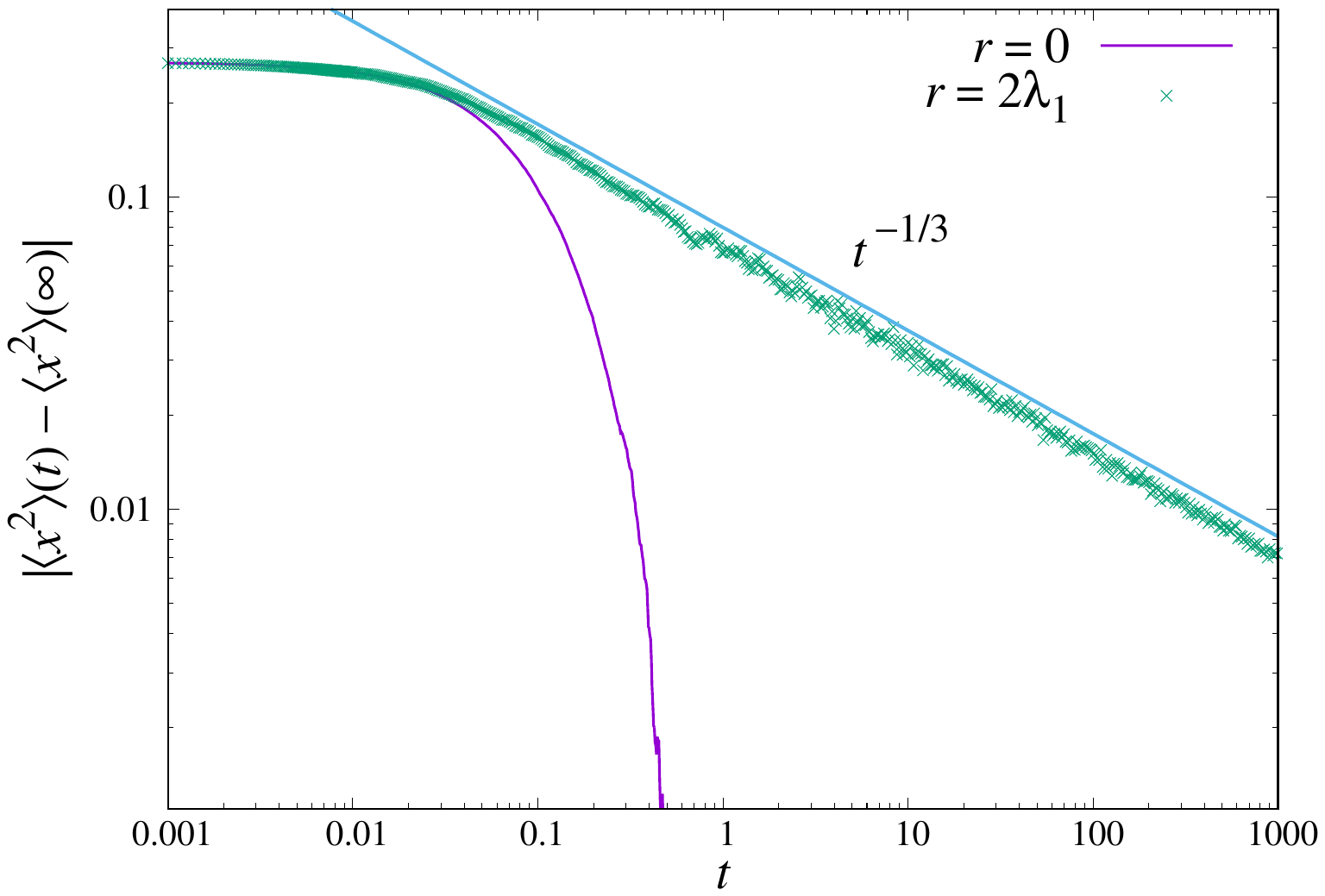}
\caption{Simulation results of the particle with memory confined in an interval  $[0,L]$ with reflecting boundary conditions. The mean square position slowly relaxes toward its asymptotic value $\langle x^2 \rangle (t\to\infty)=L^2/3$ in a power-law way (symbols), while the memory-less case $r=0$ follows an exponential behaviour (purple solid line). We have set the resetting rate to $2\lambda_1=2D\pi^2/L^2$. The scaling law $t^{-1/3}$ predicted by Eq. (\ref{box_trans.1}) is displayed as a guide to the eye. We have set $L=1$, $D=1$, $x_0=0.25$, $\Delta t=10^{-4}$ and the averages are performed over $10^4$ independent trajectories.}
\label{fig.box}
\end{figure}

\section{Numerical simulations}

To verify and illustrate these theoretical results in the different cases, we have performed numerical simulations by incorporating resetting with preferential memory in standard Brownian dynamics algorithms \cite{DPP2024}. At each discrete time $t=n\Delta t$, where $n$ a positive integer and $\Delta t$ the simulation time-step ($\Delta t\ll 1$), with probability $r \Delta t$ the particle is reset and its next position $x_{n+1}$ is given by 
\begin{equation}
x_{n+1}=x_{n'}
\end{equation}
where $n'$ is a random integer uniformly distributed in $\{0,1,2,\ldots,n\}$. With the complementary probability $1-r\Delta t$, the position evolves through the discrete Langevin equation
\begin{equation}
x_{n+1}=x_n-U'(x_n)\Delta t+\sqrt{2D\Delta t}\,\xi_{n+1},
\end{equation}
with $\xi_{n+1}$ a Gaussian random variable of zero mean and unit variance.

Instead of extracting the full distribution $P_r(x,t)$ from the simulations, we have computed the simpler mean square position of the particle $\langle x^2\rangle (t)$. If the position density obeys the generic form given by Eq. (\ref{transr.1}) and if no cancellation occurs, the second moment also relaxes toward its stationary value as
\begin{equation}
\langle x^2\rangle(t)-\langle x^2\rangle (t\to\infty)\sim t^{-\theta_1}.
\label{transvar.1}
\end{equation}
Furthermore, the single power-law decay at late time in Eq. (\ref{transvar.1}) is easier to observe if the separation between $\theta_1$ and the exponent of the second excited mode, $\theta_2$, is as large as possible. In such a case, the condition $t^{-\theta_1}\gg t^{-\theta_2}$ is fulfilled from shorter times and the contribution of the sub-leading terms in Eq. (\ref{Prxt_sol.2}) is actually negligible.

\begin{figure}
\centering
\includegraphics[width=0.8\textwidth]{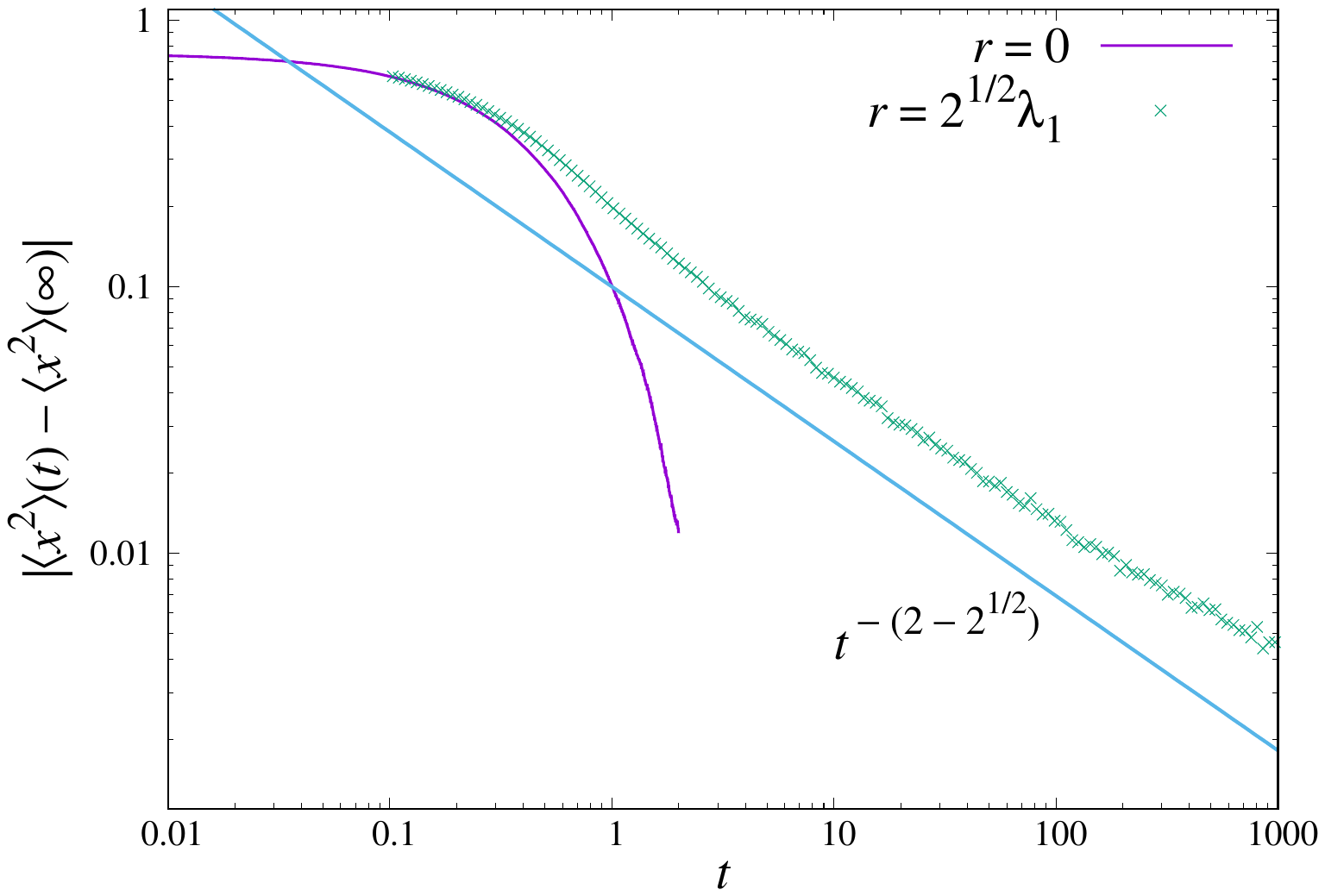}
\caption{Relaxation of the particle with memory in a harmonic potential $U(x)=\mu x^2/2$. The mean square position relaxes slowly (symbols) at finite resetting rate (here $r=\sqrt{2}\lambda_1=\sqrt{2}\mu$).  The memory-less case is shown with the purple solid line. In this case, the leading power-law decay is $t^{-\theta_2}$ with $\theta_2=2-\sqrt{2}$ (blue straight line as a guide to the eye). Here, $\langle x^2 \rangle (t\to\infty)=D/\mu$ and the stiffness is set to $\mu=1$, while $D=1$, $\Delta t=10^{-3}$, $x_0=0.5$ and the averages are preformed over $10^6$ trajectories.}
\label{fig.OU}
\end{figure}

For instance, for the particle confined in the box $[0,L]$, one has $\theta_n=\frac{D (n\pi/L)^2}{r+D (n\pi /L)^2}$ from Eq. (\ref{box_sol.2}) and the gap between the first two exponent reads
\begin{equation}
\theta_2-\theta_1=\frac{3\lambda_1r}{(r+4\lambda_1)(r+\lambda_1)}.
\end{equation}
with $\lambda_1=D\pi^2/L^2$.
Notably, fixing $\lambda_1$ and varying the resetting rate from 0 to $\infty$, the above quantity reaches a maximum at a value $r^*$ given by
\begin{equation}
r^*=2\lambda_1,
\label{rstar.box}
\end{equation}
while the corresponding exponents are $\theta_1(r^*)=1/3$ and $\theta_2(r^*)=2/3$. Simulation results are shown in Figure \ref{fig.box}, which represents the absolute value of the difference in Eq. (\ref{transvar.1}) as a function of $t$, where $\langle x^2\rangle(t\to\infty)=L^2/3$ from the uniform stationary distribution. With a resetting rate set equal to the value $r^*$ in Eq. (\ref{rstar.box}), one indeed observes a scaling law in excellent agreement with $t^{-1/3}$ over nearly five decades, thus confirming the validity of the general expression (\ref{rel.1}). In the case $r=0$, the much faster exponential decay toward the stationary state is recovered.

For the particle confined by the harmonic potential $U(x)=\mu x^2/2$, the asymptotic second moment is $\langle x^2\rangle(t\to\infty)=D/\mu$, while $\theta_n=\frac{n\mu}{r+n\mu}$. This case is a bit subtle since the eigenfunction of the first excited state is odd in $x$, given by 
\begin{equation}
\psi_{\lambda_1}(x)=\frac{2\omega^{3/4}}{(4\pi)^{1/4}}x\,e^{-\omega x^2}. 
\end{equation}
Consequently, the contribution of $\psi_{\lambda_1}(x)$ to  $\langle x^2\rangle(t)$ vanishes and the second excited state provides the leading correction for that quantity:
\begin{equation}
\langle x^2\rangle(t)-\langle x^2\rangle (t\to\infty)\sim t^{-\theta_2}.
\end{equation}
The simulation results for $r=\sqrt{2}\mu$, corresponding to $\theta_2(r)=2-\sqrt{2}=0.5857...$, are shown in Fig. \ref{fig.OU} and exhibit a good agreement with theory. The slight mismatch at late times is attributed to a lack of precision due to the rather large $\Delta t$.

\begin{figure}
\centering
\includegraphics[width=0.8\textwidth]{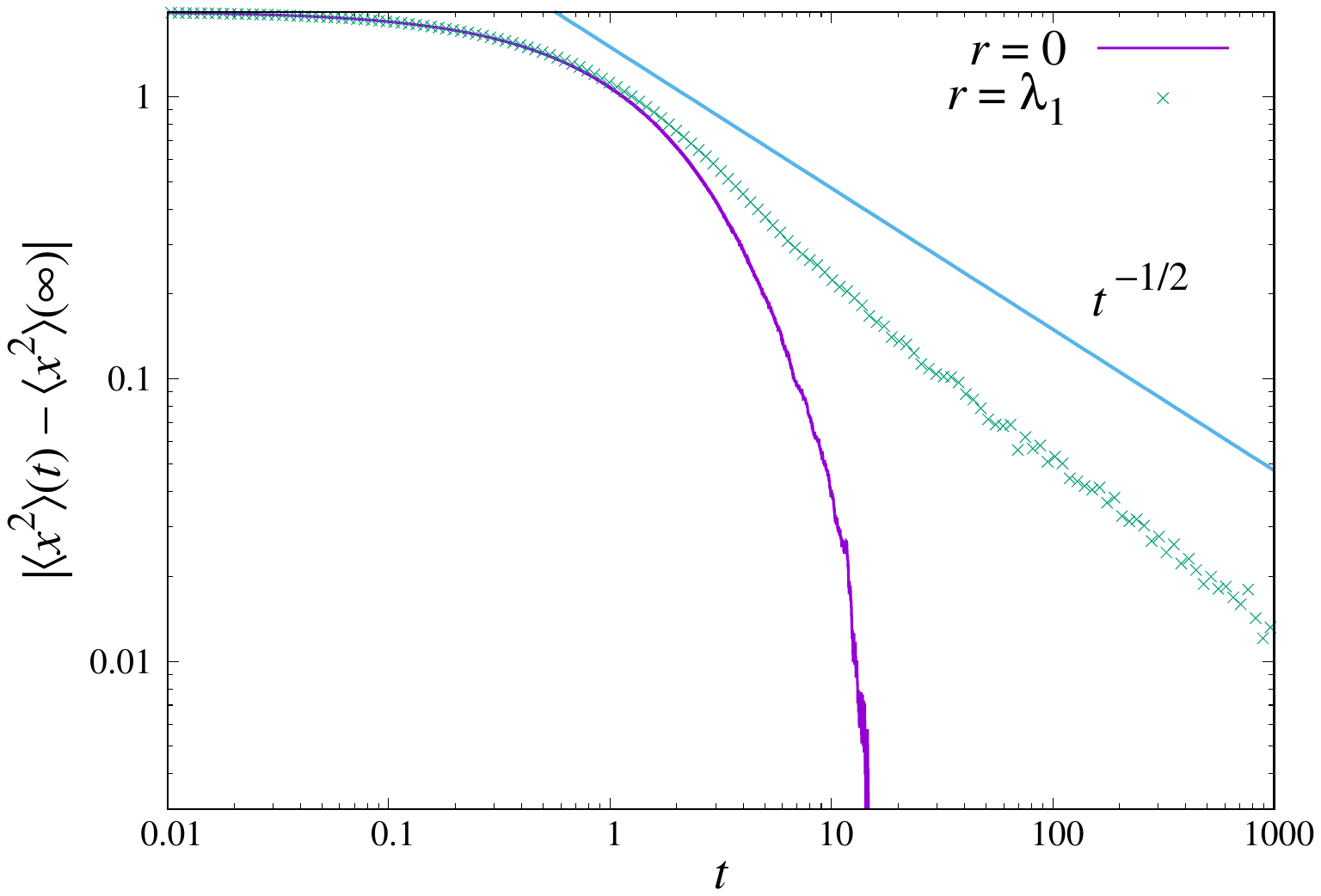}
\caption{Relaxation of the particle with memory in the potential $U(x)=F|x|$. Symbols represent the relaxation of the mean square position for a resetting rate of $r=\lambda_1=F^2/(4D)$. The case $r=0$ is represented with the purple solid line. The power-law $t^{-1/2}$ is shown as a guide to the eye. Here, $\langle x^2 \rangle (t\to\infty)=2(D/F)^2$ and the force is set to $F=1$, while $D=1$, $\Delta t=10^{-3}$, $x_0=0$.}
\label{fig.modx}
\end{figure}

Finally, in the case of the V-shaped potential $U(x)=F|x|$,  Eq. (\ref{modxr.2}) gives
\begin{equation}
\langle x^2\rangle(t)-\langle x^2\rangle (t\to\infty)\sim \left(\ln[(r+\lambda_1)t]\right)^{-3/2}\,t^{-\theta_1},
\label{transvar.2}
\end{equation}
and the second moment of the equilibrium distribution is $\langle x^2\rangle (t\to\infty)=2(D/F)^2$. Figure \ref{fig.modx} displays the mean square difference in a case where $r=\lambda_1$, i.e., where the  exponent of the leading power-law is $1/2$ according to Eq. (\ref{rel.1}). As expected, the simulation curve exhibits small deviations from a $t^{-1/2}$ law, in contrast to Fig. \ref{fig.box} of the box case that agrees perfectly with the predicted power-law. The logarithmic corrections in Eq. (\ref{transvar.2}) are difficult to detect numerically, though; we rather measure an effective exponent of roughly $0.6$ in the last two decades.

\section{Summary and Conclusion}

We have studied the relaxation of a Brownian particle with long range memory in confining potentials. The memory effects are introduced by stochastic resetting events, where the particle relocates to positions visited in the past, before it continues to diffuse in the potential from there. At large time, the particle's position tends toward a steady state given by the Gibbs-Boltzmann distribution. Contrary to confined systems subject to resetting to a fixed position \cite{P2015,ANBND2019,MBMS2020,GPKP2020,CDG2021,RR2021,ARD2022,AD2023,JBPD2023}, this stationary state is in equilibrium since local detailed balance is fulfilled. Importantly, we find that the relaxation toward the steady state is not exponential in time like in standard Brownian motion but follows an inverse power-law, for any non-zero value of the resetting rate. This sluggish decay is a consequence of the preferential character of memory in the model, where more frequently visited sites are more likely to be visited again in the future. Consequently, the particle tends to stay close to its starting point and takes a very long time to visit the accessible positions.  

The exponent that controls the slow relaxation can be obtained exactly and lies in the interval $(0,1)$. It depends on the resetting rate and on the longest relaxation time $1/\lambda_1$ of the memory-less particle through a simple general formula given by Eq. (\ref{rel.1}). In other words, knowing the asymptotic exponential relaxation of the standard Brownian particle in a potential $U(x)$ allows us to predict the anomalous decay in the presence of memory for any value of the resetting rate.  We have applied our theory to several analytically tractable potentials (box, harmonic, V-shaped), finding very good agreement with simulations. To sum up, the scale-free dynamics at large $t$ is generic and can be compared to a self-organised critical state, independent of the initial position and characterised by a non-universal exponent. 

Other models of anomalous diffusion such as continuous time random walks also exhibit a power-law relaxation toward Gibbs-Boltzmann equilibria at late times, as shown by a  fractional Fokker-Planck approach \cite{MK2000}.
The mechanism for subdiffusive motion in those processes is of very different nature and is due to a broad distribution of waiting times with infinite mean. 
The exponent $\theta_n$ of the power-law decay of the $n$-th eigenmode is also contained in the interval $(0,1)$ but turns out to be independent of $n$ and equal to the exponent associated with the waiting time distribution, or the growth of the mean square displacement. Hence, for these systems under confinement the first excited state does not dominate the relaxation and all the modes need to be considered {\em a priori}.

Our model represents a simple description of memory-based movements of an animal in its home range, i.e., a limited region of space that can be produced by an effective confining potential \cite{ML2006}, but where in addition to random motion, the living organism chooses from time to time to revisit familiar places. Empirically, space use by animals is known to be highly heterogeneous even within limited areas \cite{RPA2006,BS2014}, a feature qualitatively captured by our finding that the exploration of a small accessible region can be very slow. It would be interesting to extend this model to situations where the external potential has a local minimum and to study the crossing of the particle over a barrier.

\ack 
DB acknowledges support from CONAHCYT Ciencia de Frontera 2019 Grant No. 10872 (Mexico) and thanks the LPTMS at the Universit\'e Paris-Saclay for hospitality. SNM thanks the hospitality of KITP, Santa Barabra where this work initiated during the
workshop “Out-of-equilibrium Dynamics and Quantum Information of Many-body Systems with Long-range Interactions” (October, 2023), supported in part by the National Science
Foundation under Grant No. NSF PHY-1748958. SNM also acknowledges support from ANR Grant No. ANR-23-CE30-0020-01 EDIPS.

\end{document}